\documentclass{INTERSPEECH2023}
\usepackage{graphicx}
\usepackage{subfigure}
\usepackage{latexsym}
\usepackage{url}
\usepackage{multirow}
\usepackage{multicol}
\usepackage{makecell}
\mathchardef\mhyphen="2D

\interspeechcameraready

\title{JNV Corpus: A Corpus of Japanese Nonverbal Vocalizations with Diverse Phrases and Emotions}
\name{Detai Xin, Shinnosuke Takamichi, Hiroshi Saruwatari}
\address{
  Graduate School of Information Science and Technology, The University of Tokyo, Japan}
\email{\{detai\_xin, shinnosuke\_takamichi\}@ipc.i.u-tokyo.ac.jp}

\begin{document}

\maketitle

\begin{abstract}
We present JNV (Japanese Nonverbal Vocalizations) corpus, a corpus of Japanese nonverbal vocalizations (NVs) with diverse phrases and emotions.
Existing Japanese NV corpora lack phrase or emotion diversity, which makes it difficult to analyze NVs and support downstream tasks like emotion recognition.
We first propose a corpus-design method that contains two phases: (1) collecting NVs phrases based on crowd-sourcing; (2) recording NVs by stimulating speakers with emotional scenarios.
We then collect $420$ audio clips from $4$ speakers that cover $6$ emotions based on the proposed method.
Results of comprehensive objective and subjective experiments demonstrate that the collected NVs have high emotion recognizability and authenticity that are comparable to previous corpora of English NVs.
Additionally, we analyze the distributions of vowel types in Japanese NVs.
To our best knowledge, JNV is currently the largest Japanese NVs corpus in terms of phrase and emotion diversities.
\end{abstract}
\noindent\textbf{Index Terms}: Nonverbal vocalizations, Nonverbal emotional expressions, Corpus design, Emotion
\vspace{-3mm}
\section{Introduction}
\vspace{-1mm}
Nonverbal expressions, such as vocal, facial, and gestural expressions~\cite{tatham2004expression}, play an important role in human communication~\cite{scherer2011assessing, hall2009psychosocial}.
In human speech, nonverbal expressions are called nonverbal vocalizations (NVs), which refer to vocalizations containing no linguistic information like laughter, sobs, and screams~\cite{mehrabian2017nonverbal}.
They are relatively casual expressions and are usually used in spoken language~\cite{trouvain2012comparing}.
One of the most important functions of NVs is conveying affects~\cite{scherer1994affect, belin2008montreal}.
Such emotional NVs, which are also called affect bursts, widely exist in many cultures~\cite{sauter2010cross}.
Although NVs are ignored by most previous research on speech emotions~\cite{lima2013voices}, many works have shown the importance of NVs in emotion processing from various aspects like behavior~\cite{bestelmeyer2010auditory}, clinical treatments~\cite{dellacherie2011impaired}, human development~\cite{hunter2010effects}, and emotion recognition~\cite{xin2022exploring}.
All of these necessitate more resources and work in this field.

While there are a lot of corpora of English NVs~\cite{belin2008montreal, lima2013voices, anikin2017nonlinguistic, holz2022vivae}, the number of Japanese corpora is quite limited.
Arimoto et al. proposed OGVC~\cite{arimoto2012naturalistic}, whose utterances were collected from online game chats.
Although it includes nonverbal expressions like laughter and screams, since the designer focuses on spontaneous speech instead of emotional speech, the distribution of emotions of OGVC is unbalanced and out of control.
Other Japanese NVs corpora exist in cross-cultural researches~\cite{koeda2013cross}, Yoshie et al. studied the influence of cultural norms on the recognizability of NVs with socially disengaging emotions~\cite{yoshie2020cultural}.
They created a small Japanese NVs corpus with $144$ clips covering $9$ emotions.
While the corpus covers a wide range of emotions, since their goal is cross-cultural analysis, the phrase diversity is ignored, which results in only $2$-$3$ phrases per speaker for each emotion. 
In summary, the lack of emotion and phrase diversities make it difficult to apply current Japanese NVs corpora in real-world scenarios like emotion recognition or speech synthesis.

In this work, we present JNV (Japanese Nonverbal Vocalizations) corpus that solves the aforementioned problems.
JNV includes a total of $87$ phrases and covers $6$ emotions, which is to our best knowledge the largest Japanese NVs corpus in terms of phrase and emotion diversities.
We first propose a two-phase corpus-design method to systematically collect Japanese NVs.
The proposed method contains two phases: (1) collecting diverse phrases using crowd-sourcing; (2) recording NVs by stimulating speakers with emotional scenarios.
We then collect $420$ NVs from $4$ speakers based on the proposed method.
To validate the corpus, we conduct several subjective and objective experiments to evaluate emotion recognizability and authenticity.
Experimental results demonstrate that (1) the collected NVs can be recognized with high accuracy by both human raters and statistical models; (2) the NVs have relatively high authenticity that is comparable to previous English NVs corpus.
Furthermore, we analyze the collected phrases to show the phoneme distributions of Japanese NVs.
The contributions of this work are summarized as follows:
\begin{itemize}
    \item We propose JNV corpus, which is to our best knowledge the largest Japanese NVs corpus in terms of phrase and emotion diversities.
    \item We propose, for the first time, a scalable two-phase corpus-design method to collect Japanese NVs systematically.
    \item We conduct comprehensive subjective and objective experiments to show the NVs of JNV have high emotion recognizability and authenticity.
    \item We analyze the collected phrases of NVs and show some phonological properties of Japanese NVs.
\end{itemize}
We publicate JNV to foster future research in this field\footnote{\url{sites.google.com/site/shinnosuketakamichi/research-topics/jnv_corpus}}.
\vspace{-3mm}
\section{Proposed Method}
\vspace{-1mm}
In this section, we first describe our definition of NVs, then introduce the proposed two-phase corpus-design method.

\vspace{-3mm}
\subsection{Definition of NVs}
\vspace{-1mm}
Since there is no formalized definition of NVs, we first discuss several definitions in the literature and then give our definition.
Schröder et al. define emotional NVs as affect bursts, which refer to short and affective sounds~\cite{schroder2003experimental}.
They further extend this definition so that it includes interjective words like ``yippee".
Wharton et al. show that interjections can be used to represent emotional NVs~\cite{wharton2003interjections}.
The term interjection is also used in Japanese to represent emotional NVs~\cite{Kashino2019}.
Trouvain et al. define NVs as vocalizations in conversations that cannot be or are difficult to be transcribed into orthographical forms~\cite{trouvain2012comparing}.
In this work, we define NVs as \emph{vocalizations that can convey the emotional states of speakers but don't necessarily have their orthographical forms}.
Our definition of NVs includes not only emotional interjections but those affective sounds that cannot be transcribed. 
This is because we aim to cover as many phrases as possible and we do notice that it is difficult to transcribe some NVs.
It should be stressed that this definition also influences the recording part of the proposed method, which is explained in Section~\ref{sec:recording}.

\vspace{-3mm}
\subsection{Phase 1: phrase collection}
\vspace{-1mm}
In this phase, we aim to collect various phrases of Japanese NVs using crowd-sourcing.
We notice that emotions and NVs have a many-to-many relation, so we set two tasks for the workers:
\begin{enumerate}
    \item Emotion-to-Phrases: the workers write phrases that can express the given emotion.
    \item Phrase-to-Emotions: the workers select possible emotions that can be expressed by the given emotion.
\end{enumerate}
We use Lancers\footnote{\url{https://www.lancers.jp/}} to publish the tasks, and use the six basic emotions (anger, disgust, fear, happiness, sadness, surprise)~\cite{eckman1972universal} as the target emotions.

In the Emotion-to-Phrases task, $100$ workers are employed.
Each worker is required to give at least one phrase for each target emotion.
It results in $489$ phrases after merging the same ones.

In the Phrases-to-Emotion task, we select $80$ interjections from the Corpus of Everyday Japanese Conversation (CEJC)~\cite{koiso2022design}, which is a large conversational corpus containing massive nonverbal expressions.
We employ $100$ workers to annotate possible emotions for each interjection.
A choice of "Cannot express any emotion" is also given.
Each worker is assigned $20$ phrases hence each phrase has $25$ responses.
We then use emotions that have over $12$ votes as the labels for each phrase.
Phrases that have over $12$ "Cannot express any emotion" votes are removed.

We then combine the phrases collected from the two tasks and manually clean the data by (1) merging phrases that seem to have similar pronunciation; (2) removing phrases that are not common in Japanese; (3) removing phrases that are not consistent with our definitions of NVs.
We finally obtain 87 phrases (14/10/10/16/7/30 for anger/disgust/fear/happiness/sadness/surprise, respectively).

\vspace{-3mm}
\subsection{Phase 2: recording}
\label{sec:recording}
\vspace{-1mm}
In this phase, we record NVs using the phrases collected in phase $1$.
However, we have two problems: (1) based on our definition of NVs, there should be some NVs that are missed in phase $1$ since they can not be written in the text; (2) speakers cannot utter NVs with the designated emotion without any context.
To solve these two problems, we propose to set a phrase-free session to let the speakers utter NVs without restrictions on the phrases and use scenario-based stimuli to elicit emotions.

\vspace{-3mm}
\subsubsection{Sessions}
\vspace{-1mm}
Phrases of NVs cannot be exhaustively enumerated, thus we propose two sessions in the recording phase: (1) regular session; (2) phrase-free session.
In the regular session, we record the collected $87$ phrases for each speaker.
In the phrase-free session, we only tell speakers about the target emotion without indicating specific phrases.
Speakers are instructed to utter NVs used in their daily conversations.
Note that, we set the phrase-free session before the regular session since we expect that otherwise speakers will be influenced by the phrases given in the regular session.
However, during the recording, we also find that some speakers feel difficult to utter NVs.
We thus allow them to refer to the phrases in the regular session only in this case.

\vspace{-3mm}
\subsubsection{Stimuli}
\begin{figure}[t]
\begin{center}
\centerline{
\includegraphics[width=\columnwidth]{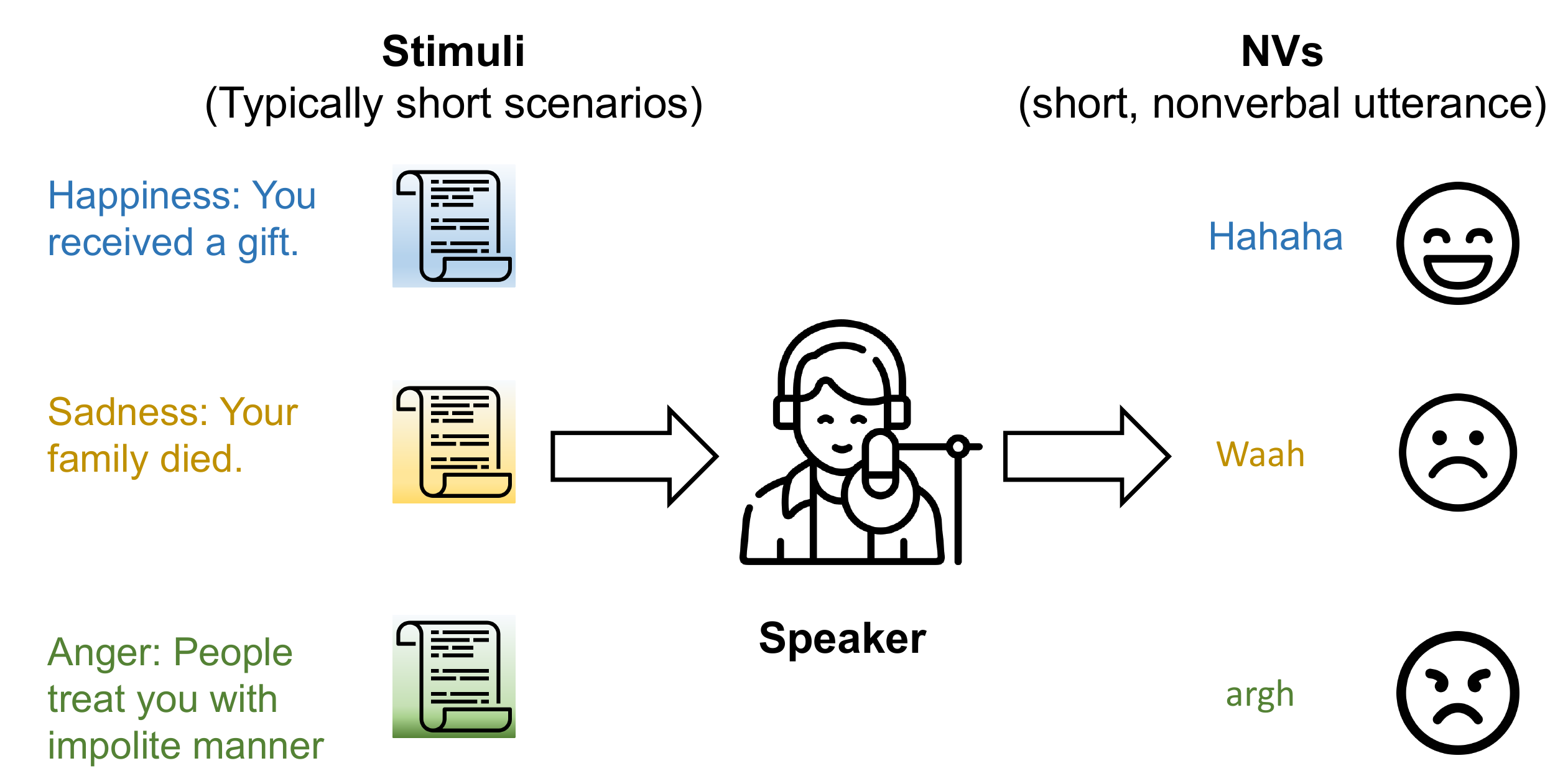}
}
\caption{The proposed recording procedure for NVs.}
\label{figure:recording}
\vspace{-11mm}
\end{center}
\end{figure}

\vspace{-1mm}
Giving stimuli to speakers for recording emotional NVs is a reliable method used in the literature~\cite{belin2008montreal}.
The stimuli can be short scenarios~\cite{lima2013voices, holz2022vivae} or videos~\cite{anikin2017nonlinguistic} that describe real-world situations containing the target emotion.
We use text-based scenarios as stimuli.
For each emotion, we make at least $3$ scenarios, which are selected and translated from the stimuli sets of previous works~\cite{belin2008montreal, lima2013voices, holz2022vivae}.
The recording procedure is illustrated in Figure~\ref{figure:recording}.
During recording, the speakers are first instructed to select a scenario from the stimuli set and imagine the situation described by it to elicit the target emotion.
Then, we ask the speakers to utter an NV to express the target emotion using the designated phrase or their preferred phrases in the phrase-free session.
We also instruct the speakers to utter NVs as naturally as possible since authenticity is quite important for emotional expressions~\cite{banziger2012introducing}.
If the recorded NV is thought to be too weak to express the target emotion, we will request the speakers a retake until the target emotion can be clearly perceived by us.
This process, though subjective, can ensure the high quality of the recorded NVs.
Note that, some speakers may be not familiar with the situations described by the stimuli, thus we also allow the speakers to imagine their personal situations as long as the target emotion can be perceived.
This is possible since in this work we don't control the emotional intensities of NVs.
Additionally, we allow the speakers to select their preferred order of emotions, but once selected, all NVs of that emotion should be recorded.
\vspace{-5mm}
\section{JNV corpus}
\vspace{-1mm}
We then make a corpus based on the proposed method.
We employ $2$ male (M1, M2) and $2$ female (F1, F2) speakers to produce the NVs.
They are native Japanese speakers and have no professional training/experience in voice acting except for one male speaker (M2).

Right before the recording, we provide speakers with an instructive document in which the recording procedure, the target emotions, and the stimuli are described.
We request the speakers to read the document roughly to get familiar with the recording.
We then invite each speaker to participate in the recording.
After an initial briefing in which we explain the goal of this recording and answer questions from the speakers about the document, we start the recording following the procedure described in the previous section.
The recording takes about two hours for each speaker with $10$ minutes rest after the first hour.

All NVs are recorded in an anechoic chamber and saved as $48$~kHz Wav files.
In the phrase-free session, the speakers are asked to utter at least $3$ NVs for each emotion, thus each speaker produces $18$ NVs in this session.
As a result, each speaker has $105$ ($87+18$) NVs, hence JNV has a total of $420$ NVs.
The total duration of JNV is $406.9$ seconds.
The average duration of high-arousal emotions like surprise ($0.8$ s) is shorter than low-arousal emotions like sad ($1.4$ s), which is intuitively consistent with the definitions of emotions.
\vspace{-3mm}
\section{Experiments}
\vspace{-1mm}
In this section, we validate and analyze the JNV corpus from various aspects.
We first conduct both objective and subjective evaluations to ensure the emotions of the NVs in JNV can be properly recognized with sufficient authenticity.
We then analyze the collected phrases to show the distributions of vowel types of Japanese NVs.

\vspace{-3mm}
\subsection{Emotion recognizability}
\vspace{-2mm}
\subsubsection{Objective classification evaluation}
\vspace{-1mm}
\label{sec:objective_classification}
We trained classification models with different combinations of features and architectures.
We used three kinds of acoustic features that are commonly used in speech emotion recognition: (1) mel-spectrograms; (2) the $6373$-dimensional ComParE set from the 2016 Computational Paralinguistics Challenge~\cite{schuller2016interspeech}; (3) the $88$-dimensional extended Geneva Minimalistic Acoustic Parameter Set (eGeMAPS)~\cite{eyben2015geneva}, in which ComParE and eGeMAPS are extracted by the openSMILE toolkit~\cite{eyben2010opensmile}.
For mel-spectrograms we used a single-layer $256$-dimensional long short-term memory (LSTM)~\cite{hochreiter1997long} to compute temporal hidden states.
The last hidden state was then fed to a linear layer with a softmax layer to get the probability distribution over the six emotions.
For the fixed-length ComParE and eGeMAPS features we used multi-layer perceptron (MLP) with the rectified nonlinear unit (ReLU) activations~\cite{Maas13rectifiernonlinearities}.
We used $5$-fold cross-validation to train all $5$ models to avoid overfitting.
Cross-entropy was used as the loss function.
The learning rate was initially set to $1e\mhyphen4$ and was halved whenever the loss value on the validation set increased.

\begin{table}[t]
    \centering
    \caption{Results of emotion recognition with $5$-fold cross-validation. \textbf{Bold} indicates the best score with $p < 0.001$.}
    \vspace{-2mm}
    \begin{tabular}{cc|c}
    \hline
    Architecture   & Feature           & ACC ($\%$)  \\
    \hline
    LSTM & mel-spec. &  $54.52 \pm 3.05$ \\
    MLP &  eGeMAPS   &  $52.60 \pm 3.93$ \\
    MLP &  ComParE   &  $\mathbf{67.12 \pm 2.60}$ \\
    \hline
    \end{tabular}
    \label{tab:obj_ser}
    \vspace{-5mm}
\end{table}
The results are demonstrated in Table~\ref{tab:obj_ser}, where classification accuracy (ACC) is used as the metric.
First, all models obtain a performance above chance ($16.67\%$), and the model using MLP and ComParE obtain the best accuracy of $67.12\%$, which shows that the emotions of the NVs in JNV can be recognized by statistical models with high accuracy.

\vspace{-3mm}
\subsubsection{Subjective forced choice evaluation}
\vspace{-1mm}
Next we conducted a forced choice task on Lancers to obtain the subjective recognition accuracy.
We asked workers to first listen to a NV, then select one emotion they thought the NV most possibly expressed.
Following previous work~\cite{lima2013voices}, we also added an extra choice "none of the above" in case the workers thought the NV could not express any of the six emotions to avoid artificially high accuracy brought by the forced choice setting~\cite{scherer2003vocal}.
A total of $42$ workers participated in this task.
Each worker was assigned $36$ NVs of which the first $6$ were dummy samples.
The dummy samples were used to enable the workers to get familiar with the task and were not counted in the final results.
We finally gathered $1260$ responses in which each NV had $3$ responses.

\begin{table}[t]
    \centering
    \caption{Distribution of responses ($\%$) in the forced choice evaluation. \textbf{Bold} indicates the best scores of each emotion.}
    \vspace{-2mm}
    \footnotesize
    \begin{tabular}{cccccccc}
    \hline
    \multirow{2}{*}{Emotion} & \multicolumn{7}{c}{Response}\\
    \cline{2-8}
    & Ang & Disg & Fear & Hap & Sad & Surp & None \\
    \hline
    Ang & $\mathbf{76.0}$ & $8.8$ & $1.0$ &$1.5$ &$2.0$ &$7.8$ &$2.9$ \\
    Disg  & $0.6$ & $\mathbf{84.0}$ & $4.9$ & $0.0$ & $2.5$ &$5.6$ &$2.5$ \\
    Fear  & $0.0$ &$14.1$ &$\mathbf{57.0}$ &$2.0$ &$0.7$ &$24.2$ &$2.0$ \\
    Hap  & $0.9$&$0.0$ &$1.3$ &$\mathbf{81.2}$ &$1.3$ &$12.5$ &$2.7$ \\
    Sad  &$2.5$ &$11.8$ &$6.7$ &$0.8$ &$\mathbf{63.9}$ &$4.2$ &$10.1$ \\
    Surp  &$2.8$ &$7.5$ &$7.5$ &$2.1$ &$0.5$ &$\mathbf{71.9}$ &$7.7$ \\
    \hline
    \end{tabular}
    \label{tab:sub_choice_emo}
    \vspace{-3mm}
\end{table}
We first calculated the distribution of responses over emotions as shown in Table~\ref{tab:sub_choice_emo}.
For all emotions, the significance is confirmed by several independent \emph{t} tests where $p \textrm{s} < 0.001$ after Bonferroni correction.
The average accuracy is $72.3\%$, which is higher than the best accuracy ($67.12$\%) obtained in the objective evaluation.
Emotions like anger ($76.0\%$), and happiness ($81.2\%$) have high recognition accuracy that is comparable to the results of Lima et al. on English where the accuracy is $78.3\%$ for anger and $85.9\%$ for pleasure (close to happiness in Russell's circumplex model~\cite{russell1980circumplex})~\cite{lima2013voices}.
Disgust has the best accuracy of $84.0\%$, which demonstrates that it is easy to be recognized.
This is also observed in several previous works on English~\cite{belin2008montreal, sauter2010perceptual, lima2013voices}.
Fear has the lowest accuracy, and $24.2\%$ of the NVs of fear were recognized as surprise.
In fact, since these two emotions are similar, such a confusion pattern is also observed in previous work on English~\cite{belin2008montreal}.
Note that, only $7.5\%$ of the NVs of surprise were recognized as fear, which we assume is because surprise as an emotion is more common than fear.
Finally, sadness has relatively low accuracy ($63.9\%$), which is usually confused with disgust ($11.8\%$) or none ($10.1\%$).
This is notably different from most previous works~\cite{belin2008montreal,lima2013voices, yoshie2020cultural} who showed high recognition accuracy on sadness.
After a careful inspection of the data, we find that the phrases of sadness in JNV contain not only sobbing but also sighs, while most previous works only contain sobbing that is easy to recognize~\cite{belin2008montreal, lima2013voices, anikin2017nonlinguistic, yoshie2020cultural}.
Actually, this difference originates from the purposes of our work and previous works.
While previous works mainly focus on scientific usages of NVs in psychology~\cite{belin2008montreal, lima2013voices, anikin2017nonlinguistic, yoshie2020cultural, holz2022vivae}, in this work we further expect JNV to have practical usages, e.g. used in expressive speech synthesis systems~\cite{kreuk2021textless}.

\begin{table}[t]
    \centering
    \caption{Emotion recognition accuracy ($\%$) of each speaker in the forced choice evaluation. \textbf{Bold} and \underline{underline} indicates the highest and lowest scores of each speaker, respectively. $*$ indicates the speaker with professional voice actor training.}
    \vspace{-2mm}
    \begin{tabular}{cccccc}
    \hline
    \multirow{2}{*}{Emotion} & \multicolumn{2}{c}{Female} & \multicolumn{2}{c}{Male} & \multirow{2}{*}{Avg.}\\
    \cline{2-5}
    & F1 & F2 & M1 & M2$^{*}$ & \\
    \hline
    Ang & $50.1$ & $86.3$ & $76.5$ & $\mathbf{90.2}$ & $76.0$ \\
    Disg & $\mathbf{86.7}$ & $87.2$ & $76.9$ & $84.6$ & $\mathbf{83.9}$ \\
    Fear  & $\underline{41.7}$ & $\underline{55.6}$ & $\underline{57.9}$ & $71.8$ & $\underline{56.7}$ \\
    Hap  & $79.0$ & $\mathbf{87.7}$ & $\mathbf{80.4}$ & $77.8$ & $81.2$ \\
    Sad  & $43.3$ & $76.7$ & $69.0$ & $66.7$ & $63.9$ \\
    Surp  & $73.7$ & $78.8$ & $69.5$ & $\underline{65.3}$ & $71.8$\\
    \hline
    Avg. & $62.6$ & $78.7$ & $71.7$ & $76.1$ & $72.3$ \\
    \hline
    \end{tabular}
    \label{tab:sub_choice_spkr}
    \vspace{-3mm}
\end{table}
Second, we calculated the accuracy of each speaker.
The result is shown in Table~\ref{tab:sub_choice_spkr}.
The performance of F1 is consistently worse than the one of others, which demonstrates that the ability to express the emotions of each speaker varies.
For the professional speaker (M2) and other speakers, although there is no obvious difference in average performance, it can be observed that M2 has significantly better performance than others in fear ($p$ $<0.001$), which shows that the professional speaker is able to well express emotions that amateur speakers are not familiar with.

To sum up, the emotions of NVs in JNV can be recognized with high accuracy by both statistical models and humans, which demonstrates the correctness and effectiveness of the proposed method.

\begin{figure}[t]
\begin{center}
\centerline{
\includegraphics[width=\columnwidth]{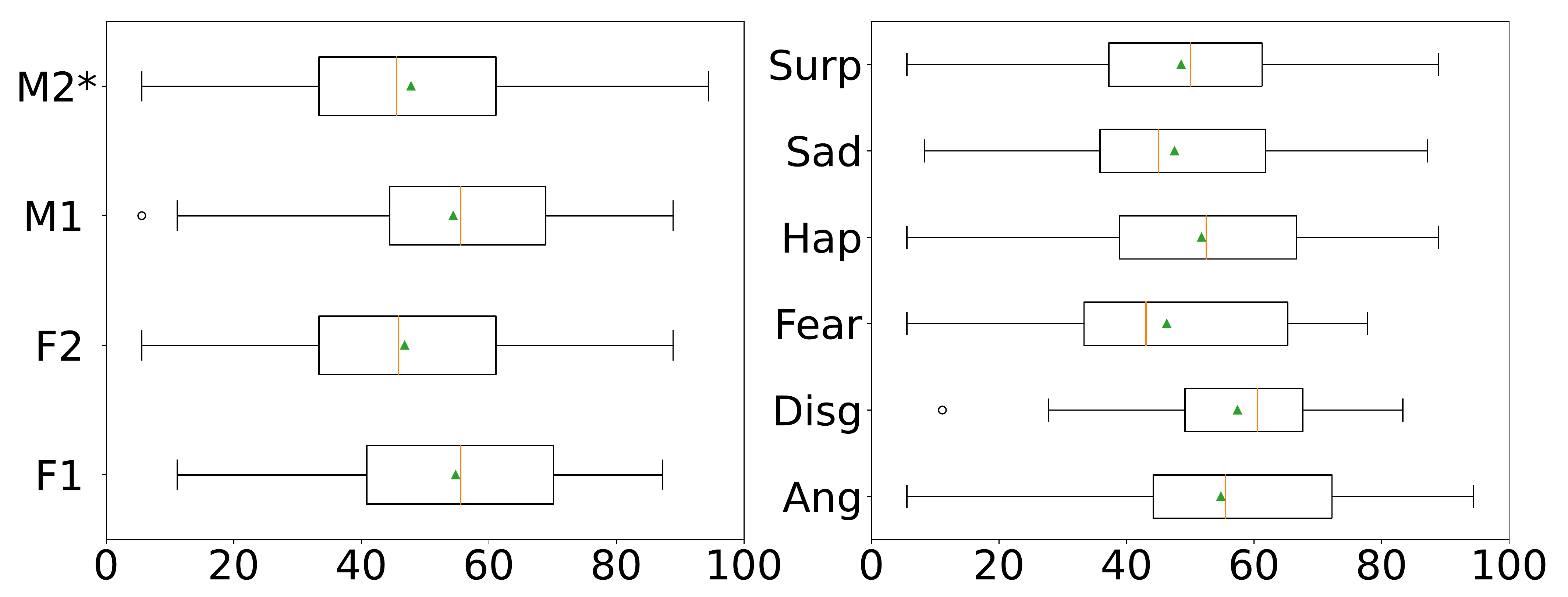}
}
\vspace{-3mm}
\caption{Left: Authenticity scores (scaled to $[0, 100]$) of speakers. $*$ indicates the speaker with professional voice training. Right: Authenticity scores (scaled to $[0, 100]$) of emotions.}
\label{figure:authenticity}
\vspace{-9mm}
\end{center}
\end{figure}

\vspace{-3mm}
\subsection{Authenticity}
\vspace{-1mm}
We then evaluated the authenticity of the NVs in JNV.
Authenticity is an important metric since it reflects to what extent the created data can be used as a replacement for real-world data, hence it is widely evaluated in previous works~\cite{lima2013voices, holz2022vivae}.
In this evaluation, for each NV the emotion and the audio were first shown to the workers, then the workers were asked to rate to what extent they thought the speaker of the NV was really experiencing the given emotion in a $7$-point Likert scale ranging from very low authenticity ($1$ point) to very high authenticity ($7$ point).
We used the same settings used in the forced choice task, i.e. the total number of workers was $42$, and each worker was assigned with $36$ NVs in which the first $6$ samples were dummy samples.
As a result, each sample had $3$ authenticity scores.
We averaged the $3$ scores for each NV and then scaled it to $[0, 100]$.

We visualize the distributions of authenticity scores over speakers and emotions using two box plots, as illustrated in Figure~\ref{figure:authenticity}.
Significant variability across speakers/emotions is confirmed by two ANOVAs ($F=5.55$, $p = 9.6e\mhyphen4$ for speakers and $F=3.28$, $p = 6.4e\mhyphen3$ for emotions).

For speakers, it can be observed that the numerical relations between them are similar to the average accuracy shown in Table~\ref{tab:sub_choice_spkr} except for the professional speaker M2, whose authenticity score is significantly lower than F2 and M1 ($p\textrm{s} < 0.01$), though his recognition accuracy ($76.1\%$) is higher than M1 ($71.7\%$) and is comparable to F2 ($78.7\%$).
This implies that the professional speaker tends to act when he produces NVs, which is arguably because he can always use his experiences in voice acting/training.

For emotions, the average authenticity score is $51.1$, which is comparable to the score obtained by Holz et al. ($4.04$ in a $7$-point scale, which is $50.7$ if scale to $[0, 100]$)~\cite{holz2022vivae}.
The scores range from $5.6$ to $94.4$.
We find many phrases that have low scores are relatively not common to express the corresponding emotions compared to those phrases that have scores, which can explain the large difference.
The numerical relations in the forced choice task shown in Table~\ref{tab:sub_choice_emo} are almost maintained in the authenticity scores of emotions, i.e. emotions that have high recognizability tend to have high authenticity scores.
This is further verified by a Pearson's correlation analysis where $r=0.87$.

In summary, the authenticity of the NVs in JNV is comparable to the one of previous works on English, which indicates the correctness and effectiveness of the proposed method.

\begin{figure}[t]
\begin{center}
\centerline{
\includegraphics[width=0.8\columnwidth]{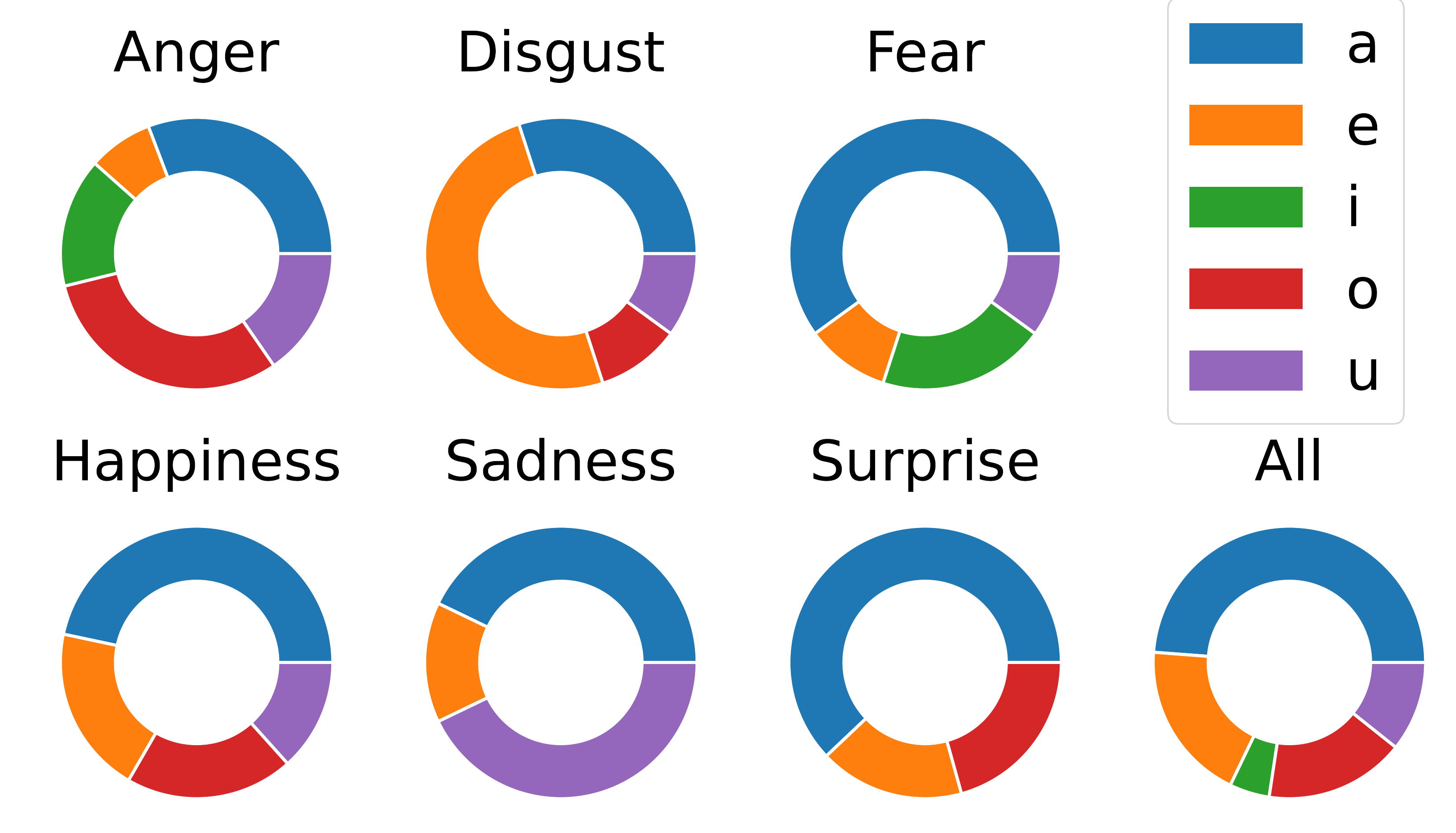}
}
\vspace{-3mm}
\caption{Distributions of predominant phonemes.}
\label{figure:phoneme}
\vspace{-11mm}
\end{center}
\end{figure}

\vspace{-3mm}
\subsection{Vowel type distributions of phrases}
\vspace{-1mm}
After technically verifying the correctness and effectiveness of the proposed method and JNV, we first analyzed the vowel type distributions of the phrases in JNV.
While Holz et al. aimed to find if there existed a vowel type that was indicative of a certain emotion~\cite{holz2022vivae}, our goal is to find general relations between vowel types and emotions in Japanese.
To ensure the phrases were common, we only selected the $87$ phrases used in the regular session.
Since each phrase could have multiple vowel types, we defined \emph{predominant vowel as the most frequent vowel in a phrase}
We then extracted the predominant vowel for each phrase by using the pykakasi package~\footnote{\url{https://github.com/miurahr/pykakasi}} to convert Hiragana into Romaji.
Note that, our method for extracting the predominant vowel is different from Holz et al. who extracted it manually from audio since they didn't collect phrases.

We visualized the distribution of vowel types over each emotion.
The results are shown in Figure~\ref{figure:phoneme}.
Each vowel type is associated with at least two emotions and each emotion is associated with at least three vowel types, which implies a many-to-many relationship between emotions and vowel types in Japanese.
This observation is consistent with English obtained by Holz et al.~\cite{holz2022vivae}.
Another observation is that ``a" is the most predominant vowel type in almost all emotions which is also found in English by Holz et al.~\cite{holz2022vivae} in which ``a" accounted for up to $50\%$ in the distributions.
Although the above two observations are common across English and Japanese, we find that ``e" is more frequent for Japanese than English.
Specifically, ``e" accounts for $19.05\%$ in our results, but only accounts for about $1\%$ in the results of Holz et al.~\cite{holz2022vivae}, which implies a possible difference between the NVs of English and Japanese.
\vspace{-3mm}
\section{Conclusions}
\vspace{-2mm}
In this work, we first presented a two-phase corpus-design method for collecting Japanese NVs systematically.
Then we proposed JNV corpus, a corpus of Japanese NVs with $420$ NVs produced by $4$ speakers covering $6$ emotions, collected by the proposed method.
Experimental results of comprehensive objective and subjective evaluations demonstrated the NVs of JNV have comparable emotion recognizability and authenticity to previous work.
We finally analyzed the collected phrases and showed the distribution of predominant vowel types in Japanese.
The differences and similarities between English and Japanese NVs were discussed by comparing our results to previous works.
To our best knowledge, JNV is currently the largest corpus of Japanese NVs in terms of phrase and emotion diversities.
In the future, we aim to synthesize NVs by using JNV to improve the expressiveness of synthetic speech.

{\normalsize
\textbf{Acknowledgements:} 
This work was supported by JST SPRING, Grant Number JPMJSP2108, JSPS KAKENHI, Grant Number JP23KJ0828.
}

\bibliographystyle{IEEEtran}
\bibliography{mybib}

\end{document}